# Mode Splitting for Efficient Plasmoinc Thin-film Solar Cell

Tong Li[1,*], Lei Dai[1], and Chun Jiang[1]

*[1]State Key Laboratory of Advanced Optical Communication Systems and Networks,
Shanghai Jiao Tong University, Shanghai, 200240, China*

*\*rilcky@sjtu.edu.cn*

**Abstract:** We propose an efficient plasmonic structure consisting of metal strips and thin-film silicon for solar energy absorption. We numerically demonstrate the absorption enhancement in symmetrical structure based on the mode coupling between the localized plasmonic mode in Ag strip pair and the excited waveguide mode in silicon slab. Then we explore the method of symmetry-breaking to excite the dark modes that can further enhance the absorption ability. We compare our structure with bare thin-film Si solar cell, and results show that the integrated quantum efficiency is improved by nearly 90% in such thin geometry. It is a promising way for the solar cell.

## References and Links

1. M. A. Green, *Solar Cells: Operating Principles, Technology and System Applications* (The University of New South Wales, Sydney, 1998).
2. S. B. Mallick, M. Agrawal, and P. Peumans, "Optimal light trapping in ultra-thin photonic crystal crystalline silicon solar cells," Opt. Express **18**, 5691-5706 (2010).
3. D. Zhou, and R. Biswas, "Photonic crystal enhanced light-trapping in thin film solar cells," Journal of Applied Physics **103**, 093102-093105 (2008).
4. T. Wakamatsu, K. Saito, Y. Sakakibara, and H. Yokoyama, "Surface Plasmon-Enhanced Photocurrent in Organic Photoelectric Cells," Jpn. J. Appl. Phys. **36** (1995).
5. O. Stenzel, A. Stendal, K. Voigtsberger, and C. von Borczyskowski, "Enhancement of the photovoltaic conversion efficiency of copper phthalocyanine thin film devices by incorporation of metal clusters," Solar Energy Materials and Solar Cells **37**, 337-348 (1995).
6. S. Pillai, and M. A. Green, "Plasmonics for photovoltaic applications," Solar Energy Materials and Solar Cells **94**, 1481-1486 (2010).
7. H. A. Atwater, and A. Polman, "Plasmonics for improved photovoltaic devices," Nat Mater **9**, 205-213 (2010).
8. K. R. Catchpole, and A. Polman, "Plasmonic solar cells," Opt. Express **16**, 21793-21800 (2008).
9. J. R. Nagel, and M. A. Scarpulla, "Enhanced absorption in optically thin solar cells by scattering from embedded dielectric nanoparticles," Opt. Express **18**, A139-A146 (2010).
10. C. Hägglund, and B. Kasemo, "Nanoparticle Plasmonics for 2D-Photovoltaics:Mechanisms, Optimization, and Limits," Opt. Express **17**, 11944-11957 (2009).
11. F.-J. Tsai, J.-Y. Wang, J.-J. Huang, Y.-W. Kiang, and C. C. Yang, "Absorption enhancement of an amorphous Si solar cell through surface plasmon-induced scattering with metal nanoparticles," Opt. Express **18**, A207-A220 (2010).
12. S. H. Lim, W. Mar, P. Matheu, D. Derkacs, and E. T. Yu, "Photocurrent spectroscopy of optical absorption

enhancement in silicon photodiodes via scattering from surface plasmon polaritons in gold nanoparticles," Journal of Applied Physics **101**, 104309-104307 (2007).
13. D. Derkacs, S. H. Lim, P. Matheu, W. Mar, and E. T. Yu, "Improved performance of amorphous silicon solar cells via scattering from surface plasmon polaritons in nearby metallic nanoparticles," Applied Physics Letters **89**, 093103-093103 (2006).
14. O. L. Muskens, J. G. Rivas, R. E. Algra, E. P. A. M. Bakkers, and A. Lagendijk, "Design of Light Scattering in Nanowire Materials for Photovoltaic Applications," Nano Letters **8**, 2638-2642 (2008).
15. V. E. Ferry, M. A. Verschuuren, H. B. T. Li, E. Verhagen, R. J. Walters, R. E. I. Schropp, H. A. Atwater, and A. Polman, "Light trapping in ultrathin plasmonic solar cells," Opt. Express **18**, A237-A245 (2010).
16. V. E. Ferry, L. A. Sweatlock, D. Pacifici, and H. A. Atwater, "Plasmonic Nanostructure Design for Efficient Light Coupling into Solar Cells," Nano Letters **8**, 4391-4397 (2008).
17. W. Bai, Q. Gan, F. Bartoli, J. Zhang, L. Cai, Y. Huang, and G. Song, "Design of plasmonic back structures for efficiency enhancement of thin-film amorphous Si solar cells," Opt. Lett. **34**, 3725-3727 (2009).
18. W. Bai, Q. Gan, G. Song, L. Chen, Z. Kafafi, and F. Bartoli, "Broadband short-range surface plasmon structures for absorption enhancement in organic photovoltaics," Opt. Express **18**, A620-A630 (2010).
19. P. N. Saeta, V. E. Ferry, D. Pacifici, J. N. Munday, and H. A. Atwater, "How much can guided modes enhance absorption in thin solar cells?," Opt. Express **17**, 20975-20990 (2009).
20. Y. A. Akimov, W. S. Koh, and K. Ostrikov, "Enhancement of optical absorption in thin-film solar cells through the excitation of higher-order nanoparticle plasmon modes," Opt. Express **17**, 10195-10205 (2009).
21. S.-J. Tsai, M. Ballarotto, D. B. Romero, W. N. Herman, H.-C. Kan, and R. J. Phaneuf, "Effect of gold nanopillar arrays on the absorption spectrum of a bulk heterojunction organic solar cell," Opt. Express **18**, A528-A535 (2010).
22. J.-Y. Wang, F.-J. Tsai, J.-J. Huang, C.-Y. Chen, N. Li, Y.-W. Kiang, and C. C. Yang, "Enhancing InGaN-based solar cell efficiency through localized surface plasmon interaction by embedding Ag nanoparticles in the absorbing layer," Opt. Express **18**, 2682-2694 (2010).
23. R. A. Pala, J. White, E. Barnard, J. Liu, and M. L. Brongersma, "Design of Plasmonic Thin-Film Solar Cells with Broadband Absorption Enhancements," Advanced Materials **21**, 3504-3509 (2009).
24. F. J. Beck, A. Polman, and K. R. Catchpole, "Tunable light trapping for solar cells using localized surface plasmons," Journal of Applied Physics **105**, 114310-114317 (2009).
25. S. Pillai, K. R. Catchpole, T. Trupke, and M. A. Green, "Surface plasmon enhanced silicon solar cells," Journal of Applied Physics **101**, 093105-093108 (2007).
26. T. L. Temple, G. D. K. Mahanama, H. S. Reehal, and D. M. Bagnall, "Influence of localized surface plasmon excitation in silver nanoparticles on the performance of silicon solar cells," Solar Energy Materials and Solar Cells **93**, 1978-1985 (2009).
27. C. Hagglund, M. Zach, G. Petersson, and B. Kasemo, "Electromagnetic coupling of light into a silicon solar cell by nanodisk plasmons," Applied Physics Letters **92**, 053110-053113 (2008).
28. F. J. Beck, S. Mokkapati, A. Polman, and K. R. Catchpole, "Asymmetry in photocurrent enhancement by plasmonic nanoparticle arrays located on the front or on the rear of solar cells," Applied Physics Letters **96**, 033113-033113 (2010).
29. S. Zhang, D. A. Genov, Y. Wang, M. Liu, and X. Zhang, "Plasmon-Induced Transparency in Metamaterials," Physical Review Letters **101**, 047401 (2008).
30. N. Liu, L. Langguth, T. Weiss, J. Kastel, M. Fleischhauer, T. Pfau, and H. Giessen, "Plasmonic analogue of electromagnetically induced transparency at the Drude damping limit," Nat Mater **8**, 758-762 (2009).
31. R. D. Kekatpure, E. S. Barnard, W. Cai, and M. L. Brongersma, "Phase-Coupled Plasmon-Induced Transparency," Physical Review Letters **104**, 243902 (2010).
32. D. M. Sullivan, *Electromagnetic simulation using the FDTD method* (IEEE Press, New York, 2000).
33. M. A. Ordal, L. L. Long, R. J. Bell, S. E. Bell, R. R. Bell, J. R. W. Alexander, and C. A. Ward, "Optical properties of the metals Al, Co, Cu, Au, Fe, Pb, Ni, Pd, Pt, Ag, Ti, and W in the infrared and far infrared," Appl. Opt. **22**, 1099-1119 (1983).
34. P. B. Johnson, and R. W. Christy, "Optical Constants of the Noble Metals," Physical Review B **6**, 4370 (1972).
35. A. Christ, O. J. F. Martin, Y. Ekinci, N. A. Gippius, and S. G. Tikhodeev, "Symmetry Breaking in a Plasmonic

Metamaterial at Optical Wavelength," Nano Letters **8**, 2171-2175 (2008).
36. K. Aydin, I. M. Pryce, and H. A. Atwater, "Symmetry breaking and strong coupling in planar optical metamaterials," Opt. Express **18**, 13407-13417 (2010).
37. F. Lasnier, and T. G. Ang, *Photovoltaic Engineering Handbook* (Adam Hilger, Bristol, 1990).
38. "ASTM standard tables for reference solar spectral irradiances: direct normaland hemispherical on 37° tilted surface. Standard No. G173-03. WestConshohocken (PA): American Society for Testing and Materials," (2003), http://www.astm.org.

## 1. Introduction

Solar cell can convert the sunlight into clean electrical power[1]. Silicon has been chosen as the material of solar cell, because of its cost-effective, earth abundance, non-toxicity and processing availability. Si solar cell will make a huge contribution to solving the energy problem in the future; however, nowadays it cannot replace traditional power, such as fuel oil, because of its high material and processing cost, and its relatively low absorbance. In recent years, many researches report on the thin-film solar cell with the thickness of a few μm or even below, much less than the crystalline silicon wafers' thickness of 180~300 μm in most of the solar cell market. The thin-film solar cell is a brilliant technology for reducing the material and processing cost of photovoltaic devices, and it also allows large-scale use of expensive scarce semiconductor materials such as indium and tellurium. But thin-film solar cell still has some limitation that the absorbance of near-bandgap light is small, especially for the indirect-bandgap semiconductor Si. People try to design efficient light-trapping schemes to enhance its absorption efficiency: one of the useful methods is to use metallic nanostructure; another choice is to utilize the photonic crystal [2-3], and so on. It was first proposed to use the unique optical properties of metallic structure (plasmonic) to boost the absorbance of photovoltaic cells [4-5] in 1990s. Those metallic structures support excitations of the conduction electrons at the interface between a metal and a dielectric, i.e. surface plasmons[6].

Surface plasmons excitations enable unparalleled light concentration and trapping, and such structures can offer at least three ways to enhance the absorbance while reducing the bulk volume and the thickness of the absorption layers[7]: Firstly, metallic nanoparticles are used as subwavelength scattering element to couple and trap free plane waves into absorption layers [8-14]. The scattered light will then acquire an angular spread in the dielectric that effectively increases the optical path length, and the shape and size of the metal nanoparticles are key factors influencing the incoupling efficiency. Secondly, a metallic back structure can couple sunlight into surface plasmon polarizations (spp) modes [15-19] at the metallic/semiconductor interface. In this geometry, the incident light is tuned 90° and the flux is absorbed in the lateral direction, which can efficiently confine and guide the light in the semiconductor layer. Thirdly, metallic nanoparticles are used as subwavelength antenna that efficiently store incident energy in the localized surface plasmon mode, which will make enhanced photocurrent owing to the plasmonic near-field coupling [20-28]. A common feature of most of the references is that they generate only one absorption-enhancement peak near some definite wavelengths to improve the solar cell efficiency, i.e. they only improve the solar energy absorption in some narrow band, and this is a limitation of the full-band solar cell absorption.

In this paper, we focus on designing a new scheme that will enhance the absorbance in much wider wavelength range, not limited in some narrow wavelength band, to improve the full-band absorption more efficiently. It is easy to acknowledge that two coupled modes with nearly resonance wavelengths and different quality factors could split the absorption peaks into more sub-peaks over wider wavelength bands, which is the core concept of the mode splitting theory. Electronic-magnetic induced transparence (EIT) [29-31] is a typical application of the mode splitting theory, and we also can apply this theory in the absorption enhancement of solar cell. We investigated a plasmonic absorption enhancement structure consisted of metal strip pairs and silicon and finite-difference time-domain (FDTD) simulation demonstrates the absorption enhancement by applying mode splitting theory. The metallic nanoparticles are designed to couple with slit cavities in the silicon slab, and the absorption peaks will be split into wider wavelength bands, and the energy absorbance will be increased. Then we introduce the symmetry-breaking in our structure to excite a dark mode coupling with the existed modes, and the absorption spectra continue to be split resulting in further enhancement of absorption. To our knowledge, no such structure has been proposed until now. We compare our structure with bare thin-film Si slab structure, and results show that the integrated quantum efficiency has been improved by nearly 90%.

## 2. Results and discussion

### 2.1. Numerical method

Our structure consists of a periodic array of Ag strip pairs and a Si slab with a period array of slits, which are both embedded in the SiO2 buffer as shown in Fig.1.(a) for 3D view. Fig.1. (b) demonstrates the cross section of this structure, and the thickness and refractive index of the $SiO_2$ buffer are 500nm and 1.5 respectively. We put the Si slab with the thickness of 120nm in the center of the buffer. An array of slits is etched on the top side of the Si slab and an array of Ag strip pairs is set on the top surface of the Si slab. We carry out numerical calculation using finite-difference time-domain (FDTD) method [32]. The complex constants of Ag and Si in our simulation are based on the experiment data [33-34]. Perfect matched layers (PML) are put on the top and bottom boundaries of our computation area and periodic boundary conditions (PBC) are set along the periodic direction. A plane wave is incident along –z direction through the absorption system. We put two monitors respectively at the top and bottom edges of the geometry to record the power of transmission (T) and reflection (R), and we use A=1-T-R to calculate the absorption power. Because the TM polarized light can drive surface plasmon resonances in metal strips, we only discuss the results of the system for the case of TM polarization.

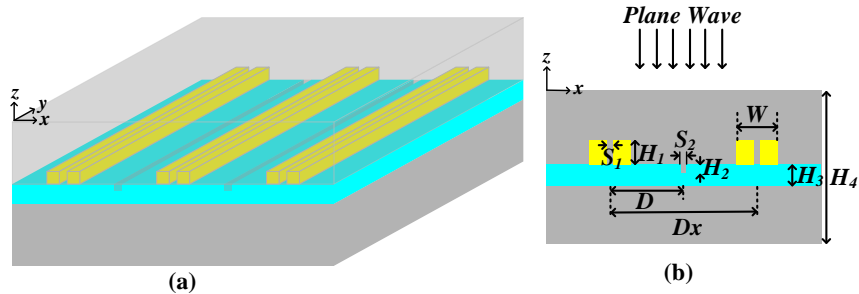

Fig. 1: Schematic of the proposed plasmonic absorption enhancement solar cell structure. (a) 3D view of our structure. (b) 2D view of the proposed structure. The geometry parameters are: S1=10nm, S2=10nm, W=100nm, H1=90nm, H2=40nm, H3=120nm, H4=500nm, Dx=350nm and D is the displacement of Ag strip pair compared with the Si slit array, which is a variable between 0nm and 175nm.

### 2.2. Absorption analysis

Firstly, we discuss the mode splitting response of our system. In fact, our proposed system consists of two sub system: One is an array of slits in Si slab and the other is an array of Ag strip pairs. The slits etched on the top surface of Si slab can be viewed as cavities which excite the coupling of free light to surface wave propagating on the corrugated surface. Fig. 2 (a) shows the absorption peak induced by the waveguide mode and the electric field intensity distribution of the peak wavelength. One Ag strip pair makes up a metal-dielectric-metal waveguide terminated by dielectric mirrors on both sides, and therefore the array of Ag strip pairs form cavities that hold Fabry-Perot mode. Fig.2 (b) shows the localized surface plasmon mode between the Ag strip pair, and the energy distributes not only in the cavity, but also on the top and bottom surfaces of the Ag strips. When combining the two sub systems together in a symmetrical structure, Fig. 2(c) shows the coupling of these two localized mode. We observe more peaks comparing to either single sub system. These resonant modes appear, due to the splitting of absorption peak of radiant states. Thus we obtain wide range absorption spectra to enhance the integrated absorbance.

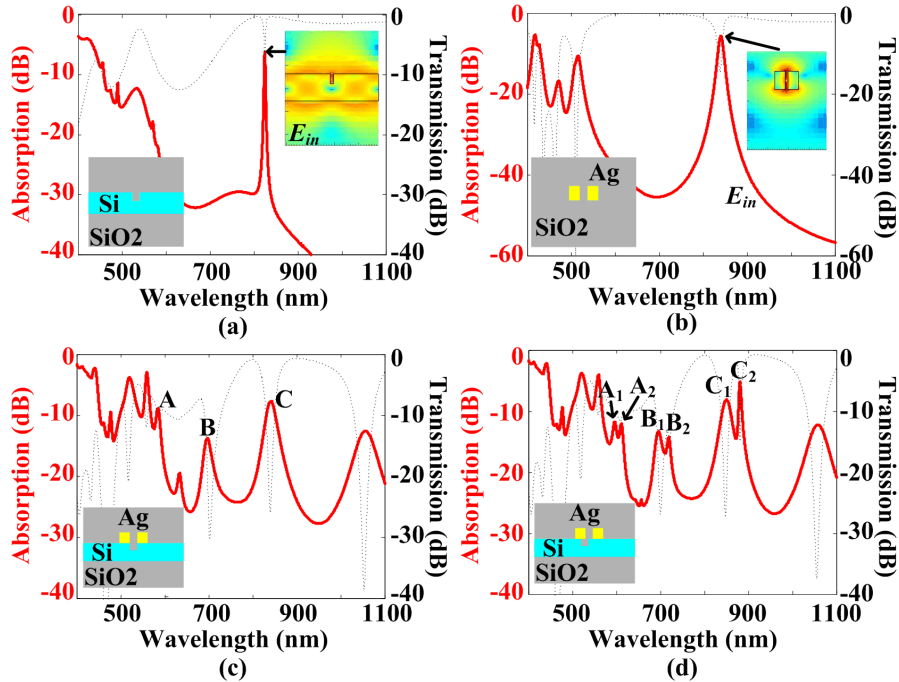

Fig.2. (a) and (b): Transmission and absorption spectra for the Si slit arrays structure and the Ag strip pair arrays structure, respectively. (c): Transmission and absorption spectra of the symmetrical coupling system with the displacement D of 0nm. (d): Transmission and absorption spectra of the asymmetrical coupling system with the displacement D of 5nm. The dark and red lines represent the transmission and absorption spectra respectively, and the insets are the schematic of corresponding structures.

In recent years, the concept of symmetry-breaking has been applied on metallic nanostructures [30, 35-36]. Specifically, the stacked plasmonic structure [25] with symmetry-breaking could excite the coupling of both the dark modes and the dipolar modes, which exhibits EIT-like phenomenon with absorption spectra splitting. Inspired by this idea, for our structure we also could employ the concept of symmetry-breaking to broaden the absorption wavelength range as shown in Fig.2 (d), in which the Ag strip pair moves away from the symmetrical center of Si slit. We can observe that there are several absorption peak pairs for the asymmetrical structure which are split from the absorption peak of symmetrical structure due to the symmetry-breaking. In order to reveal the feature of these new excited modes, Fig.3 shows the electric field distributions of these excited modes for the symmetrical structure with three locations A, B, C and for the asymmetrical structure with six locations A1, A2, B1, B2, C1, C2. We have a check in the upper row in Fig. 3: A is 584.3nm in the symmetrical system; A1 and A2 are 596nm and 611.2nm respectively in the asymmetrical system. Through a carefully check, we notice that the field pattern of A is symmetrical distribution around the Ag strip pair, but for the asymmetrical system the field pattern in the foot of the Ag strip pair shows novel feature where the field energy is focused mainly on the right and left foot of Ag strip pair at location A1 and A2 respectively, so it is known that the excited modes at locations A1 and A2 for the asymmetrical structure are split from the excited mode of symmetrical structure. The exciting mechanism of another absorption peak pairs (B1, B2; and C1, C2) for the asymmetrical structure also can be analyzed similarly from the figures in Fig.3.

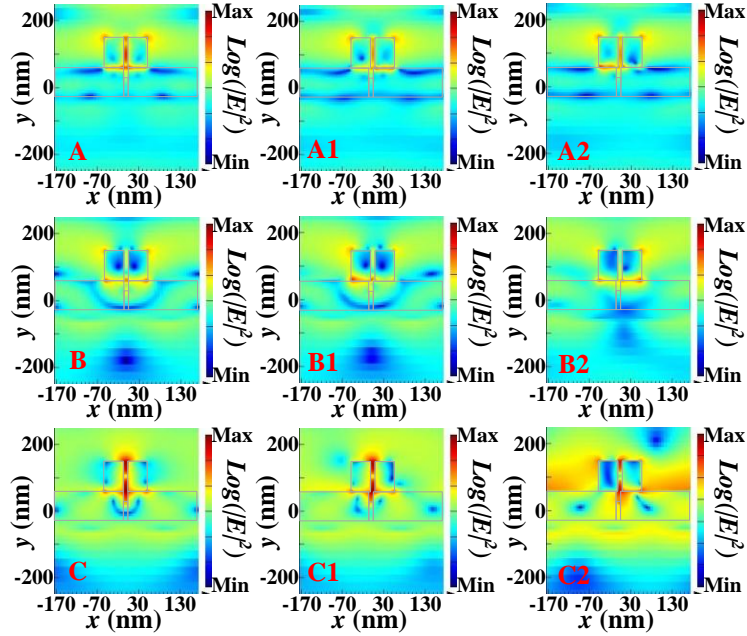

Fig. 3: The distribution of electric field intensity, which are corresponding to the A, B, C, A1, A2, B1, B2, C1, C2 point in Fig 2 (c)(d), respectively. A is 584.6nm, A1 is 596nm, A2 is 611.2nm, B is 696.5nm, B1 is 695.3nm, B2 is 719nm, C is 840.8nm, C1 is 850.1nm, C2 is 880.8nm.

## 2.3. Solar cell design

Based on the forgoing discussion of our coupling system, we can conclude that the absorption ability of coupling system can be enhanced simultaneously by constructing the interaction between different excited-modes and breaking the symmetry of system, so we can explore the proposed structure to design efficient thin-film Si solar cell. Simply, when the absorption peaks are split into more sub-peaks, the structure can hold more absorption energy over a wide wavelength range which can enhance the full-band integrated absorption. In order to evaluate the efficiency of solar cell accurately, in the FDTD simulation we interpolate two monitors at the top and bottom surfaces of the Si slab respectively to record the transmission energy of the two surfaces, and we use the difference value of their results to get the absorption energy of the solar cell $P_{abs}(\lambda)$. Meanwhile, we calculate the absorption energy $P_{abs-ref}(\lambda)$ of a bare Si solar cell as a reference. Thus, we can define the absorption enhancement G to demonstrate

the enhancement ability of the plasmonic Si solar cell:

$$G(\lambda) = \frac{P_{abs}(\lambda)}{P_{abs-ref}(\lambda)} \quad (1)$$

Fig.4 (a) shows the absorption enhancement for the symmetrical and asymmetrical structures. The parameter D in the figure is defined as the displacement of the Ag strip pairs from the center of the Si slit. From the results we can observe obviously that the value of absorption enhancement for the plasmonic Si solar cell with symmetry-breaking is higher than the value of the symmetrical structure, for example, the largest absorption enhancement is 28dB for the asymmetrical structure with displacement D of 50nm. Meanwhile, the large absorption enhancement for the asymmetrical structure distributes over wider wavelength range compared with the symmetrical case. In order to highlight the advantage of our structure we next study the absorption enhancement of traditional plasmonic solar cell as shown in Fig.4 (b), where the structure consists of one Ag strip on the top of Si slab in a period. Results indicate that the absorption enhancement of our structure has more and higher peaks in the whole band. Thus, we can conclude that our structure based on the symmetry-breaking can be explored to design efficient thin-film solar cell.

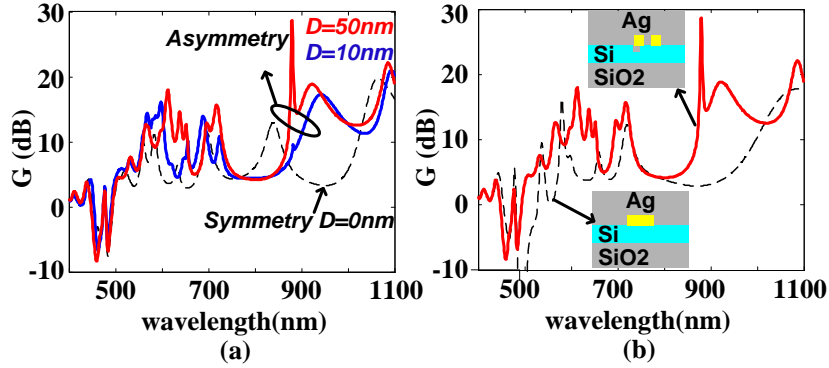

Fig. 4: Illustration of absorption enhancement. (a) Absorption enhancement of the symmetrical structure and the asymmetrical structure; (b) absorption enhancement comparison of our structure to no-slit situation, the inset shows the schematic of the structures.

Finally, we define the integrated quantum efficiency (IQE) and short circuit density (J) [37] to evaluate our solar cell more intuitively:

$$IQE = \frac{\int \frac{\lambda}{hc} QE(\lambda) I_{AM1.5}(\lambda) d\lambda}{\int \frac{\lambda}{hc} I_{AM1.5}(\lambda) d\lambda} \quad (2)$$

$$J = e \int \frac{\lambda}{hc} QE(\lambda) I_{AM1.5}(\lambda) d\lambda \quad (3)$$

where $h$ is Plank's constant, $c$ is the speed of light in the free space, e is the charge of an electron, $I_{AM1.5}$ is standard AM 1.5 solar spectrum[38], and $QE(\lambda)$ is the quantum efficiency defined by:

$$QE(\lambda) = \frac{P_{abs}(\lambda)}{P_{in}(\lambda)} \quad (4)$$

$P_{abs}(\lambda)$ and $P_{in}(\lambda)$ are the power of the absorption and incident light within the solar cell

respectively.

The results for the bare Si structure are: $IQE_{ref} = 0.0720$ and the $J_{ref} = 30.3658 mA/cm^2$. The results for the traditional structure in Fig. 4(b) are: $IQE = 0.0995$, $J = 41.9483 mA/cm^2$ and the results for our proposed structure in symmetrical situation are $IQE = 0.1102$ and $J = 46.4573 mA/cm^2$. The traditional structure has IQE enhancement by 38.2%; comparatively speaking, our design of symmetrical structure has IQE enhancement by 53.1%; our asymmetrical structures have an even higher improvement of IQE and the short circuit current, as shown in Fig. 5. From equation (2) and (3), IQE and J only have a constant ratio difference, so in Fig. 5 IQE enhancement and J share the same curve shape, but with different values. The asymmetry induced mode splitting can more effectively increase IQE and J, and all the results in Fig. 5 are greater than that of the symmetrical structure (D = 0nm). Especially in the case of D = 50nm, its $IQE = 0.1359$ and $J = 57.2917 mA/cm^2$, that is nearly 90 percent more than that of the reference bare Si structure. The results show the solar absorption improvement effect of our proposed plasmonic structure based on mode splitting.

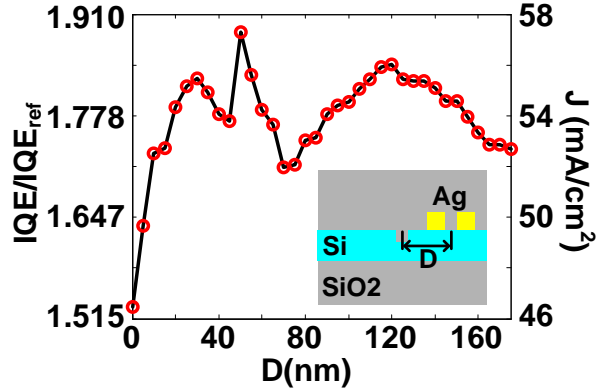

Fig. 5: The enhancement of integrated quantum efficiency IQE and the short circuit current J of the asymmetrical system, as a variable of Ag strip pairs displacement D. The inset shows the schematic and illustrates the displacement D of Ag strip pairs.

## 3. Conclusion

In summary, we propose a new structure to enhance the optical absorption in thin-film Si solar cell by applying mode splitting theory. Two mechanisms are demonstrated to successfully improve the absorbance of the solar energy: Firstly, the Ag strip pairs and the Si slab with period slit could respectively excite the localized plasmonic mode and the surface waveguide mode, which interact with each other to split the absorption spectra into wider wavelength range, and greatly increase the absorption efficiency; Secondly, the symmetry-breaking introduced in our proposed structure could excite a new dark mode, which is hidden in the symmetrical structure, to couple with the existed modes, and the coupling could further enhance the absorption of full solar spectrum. The proposed structure with a thickness of only 120nm can increase the absorption efficiency by nearly 90% at most comparing with the bare Si slab structure. For it high efficiency and simple configuration, the new scheme provides a promising way for thin-film solar cell.

**Acknowledgements**

This work was supported by National Natural Science Foundation of China (Grant No. 60672017) and sponsored by Shanghai Pujiang Program.